\documentstyle[elsart12]{elsart}
\input psfig.tex
\begin{document}

\begin{frontmatter}

\title{Differential Cross Section for $W$
       Boson Production as a Function of Transverse Momentum 
       in {\mbox{$p\bar p$}}\ Collisions at 
       {\mbox{$\sqrt{s}$ =\ 1.8\ TeV}} }
%
\begin{flushleft}
\author{ \small         
V.M.~Abazov,$^{23}$                                                           
B.~Abbott,$^{58}$                                                             
A.~Abdesselam,$^{11}$                                                         
M.~Abolins,$^{51}$                                                            
V.~Abramov,$^{26}$                                                
B.S.~Acharya,$^{17}$}
\author{ \small                                                      
D.L.~Adams,$^{60}$                                                            
M.~Adams,$^{38}$                                                              
S.N.~Ahmed,$^{21}$                                                            
G.D.~Alexeev,$^{23}$
% ...
}                    
\end{flushleft}
                                                                            
\vskip 0.70cm                                                                 
\centerline{(D\O\ Collaboration)}                                             
\vskip 0.90cm                                                                 

{\small

\begin{center}
% ...
\address{$^{11}$Laboratoire de l'Acc\'el\'erateur Lin\'eaire,              
                  IN2P3-CNRS, Orsay, France}                                  
% ...
\address{$^{17}$Tata Institute of Fundamental Research, Mumbai, India}     
% ...
\address{$^{21}$University of Nijmegen/NIKHEF, Nijmegen, The               
                  Netherlands}                                                
% ...
\address{$^{23}$Joint Institute for Nuclear Research, Dubna, Russia}       
% ...
\address{$^{26}$Institute for High Energy Physics, Protvino, Russia}       
% ...
\address{$^{38}$University of Illinois at Chicago, Chicago,                
                   Illinois 60607}                                            
% ...
\address{$^{51}$Michigan State University, East Lansing, Michigan 48824}   
% ...
\address{$^{58}$University of Oklahoma, Norman, Oklahoma 73019}            
% ...
\address{$^{60}$University of Texas, Arlington, Texas 76019}               
% ...

\end{center}
}
                                                   
\begin{abstract}
We report a measurement of the differential cross section 
for $W$ boson production as a function of its transverse momentum 
in proton-antiproton collisions at {\mbox{$\sqrt{s}$ =\ 1.8\ TeV}}. 
The data were collected by the D\O\ experiment at the Fermilab Tevatron Collider
during 1994--1995 and correspond to an integrated luminosity of $85$~pb$^{-1}$.
The results are in good agreement with quantum chromodynamics
over the entire range of transverse momentum.
\end{abstract}

\end {frontmatter}
Measurement of the differential cross section for $W$ boson production 
provides an 
important test of our understanding of quantum chromodynamics (QCD).
Its implications range from impact on the precision determination of 
the $W$ boson mass to background estimates for new physics phenomena.
Data from the production of $W$ and $Z$ bosons at hadron colliders 
also provide bounds on parametrizations used to describe the 
nonperturbative regime of QCD processes. 

The production of $W$ bosons at the Fermilab Tevatron proton-antiproton 
collider proceeds predominantly via quark-antiquark annihilation.
In the QCD description of the production mechanism,
the $W$ boson acquires transverse momentum by recoiling against additional 
gluons or quarks,
which at first order originate from the processes $q q' \rightarrow W g$
and $q g \rightarrow W q'$.
When the transverse momentum ($p_T^W$) and the invariant mass ($M_W$) 
of the $W$ boson are of the same order, 
the production rate can be calculated perturbatively order by order in the
strong coupling constant $\alpha_s$~\cite{Arnold:1989dp}.
For $p_T^W \ll M_W$, the calculation is dominated by large logarithms 
$\approx\alpha_s\ln(M_W/p_T^W)^2$, which are related 
to the presence of soft and collinear gluon radiation.
Therefore, at sufficiently small $p_T^W$, fixed-order perturbation theory 
breaks down and the logarithms must be resummed~\cite{Collins:1985kg}.
The resummation 
can be carried out in transverse momentum ($p_T$) space~\cite{Dokshitzer:1978yd} 
or in impact parameter ($b$) space~\cite{Parisi:1979se} 
via a Fourier transform.
Differences between the two formalisms are discussed in Ref.~\cite{Arnold:1991yk}.

Although resummation extends the perturbative calculation to lower values 
of $p_T^W$, a more fundamental barrier is encountered when
$p_T^W$ approaches $\Lambda_{\mbox{\scriptsize QCD}}$, 
the scale characterizing QCD processes. 
The strong coupling constant $\alpha_s$ becomes large 
and the perturbative calculation is no longer reliable.
The problem is circumvented by using a cutoff value
and by introducing an additional function that parametrizes the 
nonperturbative effects~\cite{Ladinsky:1994zn,Ellis:1998ii}.
The specific form of this function and the particular choices for the
nonperturbative parameters have to
be adjusted to give the best possible description of the data.

We report a new measurement~\cite{Mostafa:2000}
of the inclusive differential cross section for $W$ boson production in
the electron channel as a function of transverse momentum.
We use $85$~pb$^{-1}$ of data recorded with the  D\O\ detector 
during the 1994--1995 run of the Fermilab
Tevatron $p\overline{p}$ collider.
We have a ten-fold increase in the number of $W$ boson candidates
with respect to our previous measurement~\cite{Abbott:1998jy},
reflecting the larger data set and an increase in electron rapidity coverage. 
An improved electron identification technique reduces the background 
for central rapidities and high $p_T^W$ by a factor of 
five compared to Ref.~\cite{Abbott:1998jy},
and keeps the background contamination at a low level for large rapidities.
Furthermore, corrections for detector 
resolution now enable direct comparison with theory. 

Electrons are detected in an electromagnetic (EM) calorimeter 
which has a fractional energy resolution of 
$\approx 15\%/\sqrt{E(\mbox{GeV})}$ and a segmentation of
$\Delta\eta\times\Delta\phi = 0.1\times 0.1$ in pseudorapidity 
($\eta$) 
and azimuth ($\phi$).
The D\O\ detector and the methods used to select $W\rightarrow e\nu$ events
are discussed in detail in Refs.~\cite{Abachi:1994em} and \cite{Abbott:2000aj}
respectively.
Below, we briefly describe the main selection requirements.

Electron candidates are identified as isolated clusters of energy in 
the EM calorimeter that have a matching track in one of the drift chambers. 
In event reconstruction,
electron identification is based on a likelihood technique~\cite{Abbott:1998dc}.  
The electron likelihood is constructed from:
(\textit{i}) a $\chi^2$ based on a covariance matrix that determines the consistency 
of the cluster in the calorimeter with the expected shape of an electron shower,
(\textit{ii}) the ``electromagnetic energy fraction,'' defined as the ratio of the 
portion of the energy of the cluster found in the EM calorimeter to its total energy,
(\textit{iii}) a measure of the consistency between the track position and 
the centroid of the cluster, and
(\textit{iv}) the ionization energy loss along the track.
To a good approximation, 
these four variables are independent of each other. 
Electron candidates are accepted either in the central region, 
$|\eta_{\mbox{\scriptsize det}}|\le1.1$, or in the forward region,
$1.5\le|\eta_{\mbox{\scriptsize det}}|\le2.5$, 
where $\eta_{\mbox{\scriptsize det}}$ refers to the value of $\eta$ obtained by assuming
that the particle originates from the geometrical center of the D\O\ detector.

Neutrinos do not interact in the detector
and thereby create an apparent momentum imbalance. 
For each event, the missing transverse energy 
(${\hbox{$E$\kern-0.6em\lower-.1ex\hbox{/}}}_T$),
obtained from the vectorial sum of the transverse energy of all calorimeter cells,
is attributed to the neutrino.

Candidates for the $W\rightarrow e\nu$ event sample are required to have an electron with 
$E_T > 25\ \mbox{GeV}$ and 
${\hbox{$E$\kern-0.6em\lower-.1ex\hbox{/}}}_T > 25\ \mbox{GeV}$.
Additionally, 
events containing a second electron are rejected if the dielectron invariant mass 
$M_{ee}$ is close to that of the $Z$ boson 
($75\ \mbox{GeV}/c^2 < M_{ee} < 105\ \mbox{GeV}/c^2$).
A total of 50,486 events passes this selection. 

A major source of background 
stems from jets and direct photons passing our electron selection criteria.
A multijet event can be misinterpreted as a $W\rightarrow e\nu$ decay 
if one of the jets mimics an electron and there is sufficient
mismeasurement of energy to produce significant
${\hbox{$E$\kern-0.6em\lower-.1ex\hbox{/}}}_T$.
The fraction of background events due to multijet, $b$ quark, and direct-photon sources,
also referred to as QCD background, is calculated by studying the electron likelihood
in both a background sample and a signal sample,
as described in Ref.~\cite{Abbott:2000tt}.
The total QCD background in the data sample is $2\%$;
its shape is determined by repeating the background calculation for each $p_T^W$ bin.

Other sources of background in the $W\rightarrow e\nu$ sample are
$W\rightarrow \tau\nu$, $Z\rightarrow e e$, and $t\bar{t}$ events.
The process $W\rightarrow \tau\nu\rightarrow e\nu\nu\nu$ 
is indistinguishable from the signal on an event-by-event basis.
To estimate this background,
$W\rightarrow \tau\nu$ events are generated with the same $W$
boson production and decay model used in the calculation of the acceptance (see below),
and the $\tau$ leptons are forced to decay to electrons.
Since the three-body decay of the $\tau$ leads to a very soft electron $p_T$ spectrum 
compared to that from $W\rightarrow e \nu$ events,
the kinematic requirements keep this background to a moderate $2\%$.
This is accounted for
by making a correction to the acceptance for $W$ bosons~\cite{Abbott:2000tt}.
A $Z\rightarrow e e$ event can be misidentified when one of the two electrons 
escapes detection or is
poorly reconstructed in the detector and thereby simulates the presence of a neutrino.
This background ($0.5\%$) is estimated by applying the selection criteria to a sample of
Monte Carlo $Z\rightarrow e e$ events that were generated with 
{\sc isajet}~\cite{Baer:1999sp},
processed through a {\sc geant}-based~\cite{Carminati:1993} 
simulation of the D\O\ detector, and overlaid with events from
random $p\bar{p}$ crossings that follow the luminosity profile of the data.
The background from top quarks decaying to $W$ bosons 
($0.1\%$) is estimated using {\sc herwig}~\cite{Marchesini:1992ch} 
Monte Carlo $t\bar{t}$ events and {\sc geant} detector simulation.

Trigger and selection efficiencies are determined using 
$Z\rightarrow e e$ data in which one of the electrons 
satisfies the trigger and selection criteria,
and the second electron provides an unbiased sample 
to measure the efficiencies. 
Due to the limited statistics of the $Z\rightarrow e e$ data sample,
we determine the shape of the efficiency 
as a function of transverse momentum 
using $Z\rightarrow e e$ events generated with {\sc herwig}, 
processed with a {\sc geant} detector simulation,
and overlaid with randomly selected minimum-bias $p\overline{p}$ collisions.
This procedure models the effects of the underlying event 
and of jet activity on the selection of electrons.
The efficiency for both the electron identification 
and the trigger requirements is $(55.3\pm 2.2) \%$.

The data are corrected for kinematic and geometric acceptance and detector 
resolution, as determined 
from a Monte Carlo program
originally developed for measuring the mass of the $W$ boson~\cite{Abbott:1998ww}. 
The method is described in detail in Ref.~\cite{Abbott:2000wk}.
The program first generates $W$ bosons with $\eta$ and $p_T^W$ values chosen randomly 
from a double differential cross section
$d^2\sigma/dp_T^W d\eta$ provided as input.
The response of the detector and the effects of 
geometric and kinematic selection criteria
are introduced at the next stage.
For the present analysis, the input 
$d^2\sigma/dp_T^W d\eta$ distribution is obtained using the iterative unfolding
method described in Ref.~\cite{Lindemann:1995ut}.
The uncertainty due to this input distribution is evaluated by using an
initial distribution uniform in $p_T^W$ and $\eta$.
The systematic smearing uncertainty is determined by varying 
the detector resolution parameters by $\pm 1$ standard deviation from the nominal values.
The total correction for kinematic and geometric acceptance and detector
resolution for $W\rightarrow e\nu$ events is $(47.6\pm0.3) \%$.

\begin{figure}[p]
\centerline{\psfig{figure=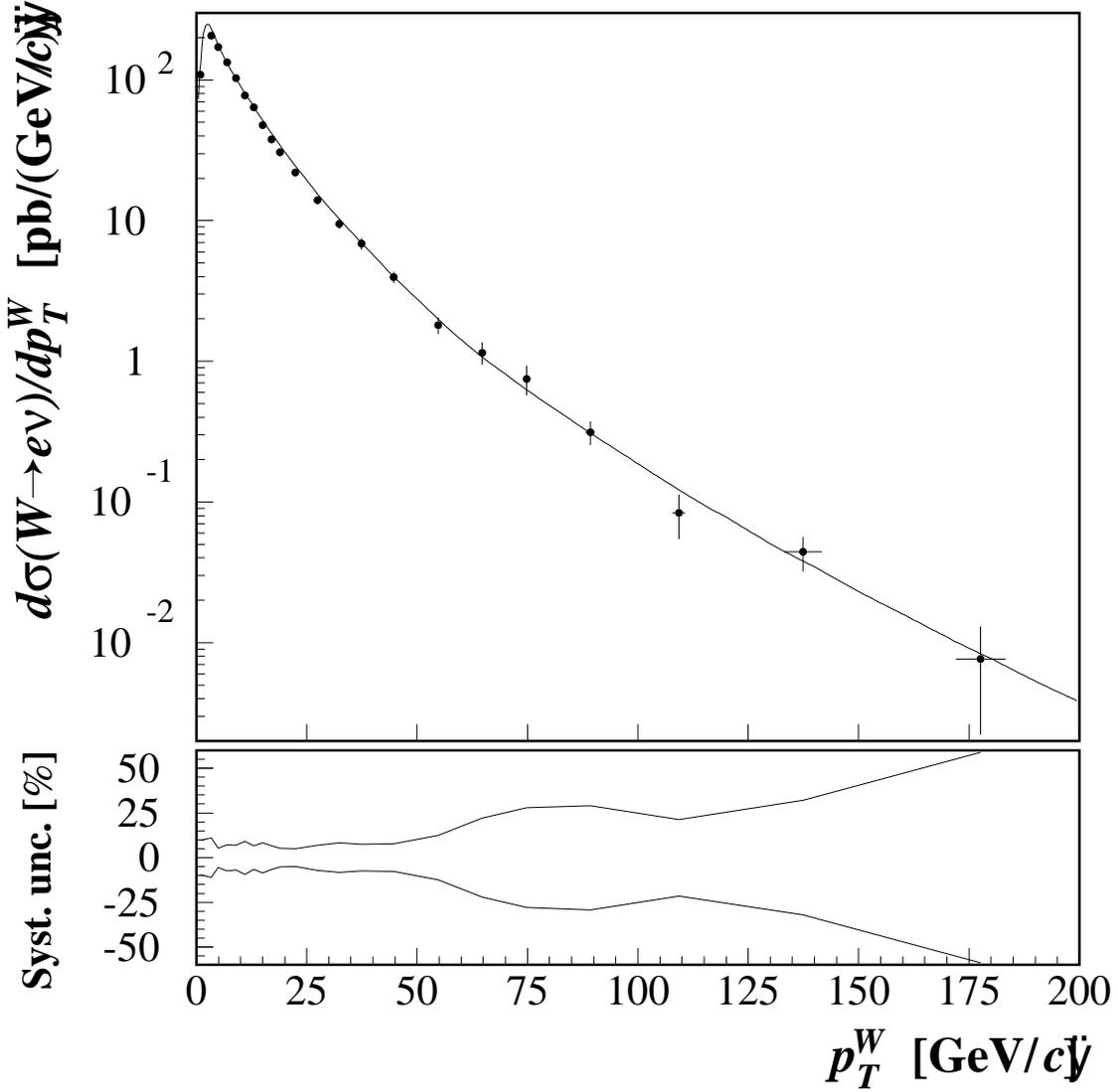,width=15cm}}
\caption{Differential cross section for $W\rightarrow e\nu$ production.
The solid line is the theoretical prediction of Ref.~\protect\cite{Ladinsky:1994zn}. 
Data points show only statistical uncertainties. 
The fractional systematic uncertainty, shown as the band in the lower plot, 
does not include an overall $4.4 \%$ normalization uncertainty in integrated luminosity.}
\label{fig:ptw}
\end{figure}

The results for $d\sigma(W\rightarrow e\nu)/dp_T^W$, 
corrected for detector acceptance and resolution, are shown in Table~\ref{tab:ptw} 
and plotted in Fig.~\ref{fig:ptw}, where the data are compared
to the combined QCD perturbative and resummed calculation in $b$-space, 
computed with published values of the nonperturbative parameters~\cite{Ladinsky:1994zn}.
The error bars on the data points correspond to their statistical uncertainties.
The fractional systematic uncertainty is shown as a band in the lower portion of the plot.
The largest contributions to the systematic error are from uncertainties in 
the hadronic energy scale and resolution, 
the selection efficiency, and the background (in the high $p_T^W$ region).
An additional normalization uncertainty of $\pm 4.4 \%$ from the integrated
luminosity is not included in any of the plots nor in the table. 
The data are normalized to the measured $W\rightarrow e\nu$ cross section 
($2310\ \mbox{pb}$~\cite{Abbott:2000tt}). 
The points are plotted at the values of $p_T^W$ where the predicted function
equals its mean over the bin~\cite{Lafferty:1995cj}.

\begin{figure}[p]
\centerline{\psfig{figure=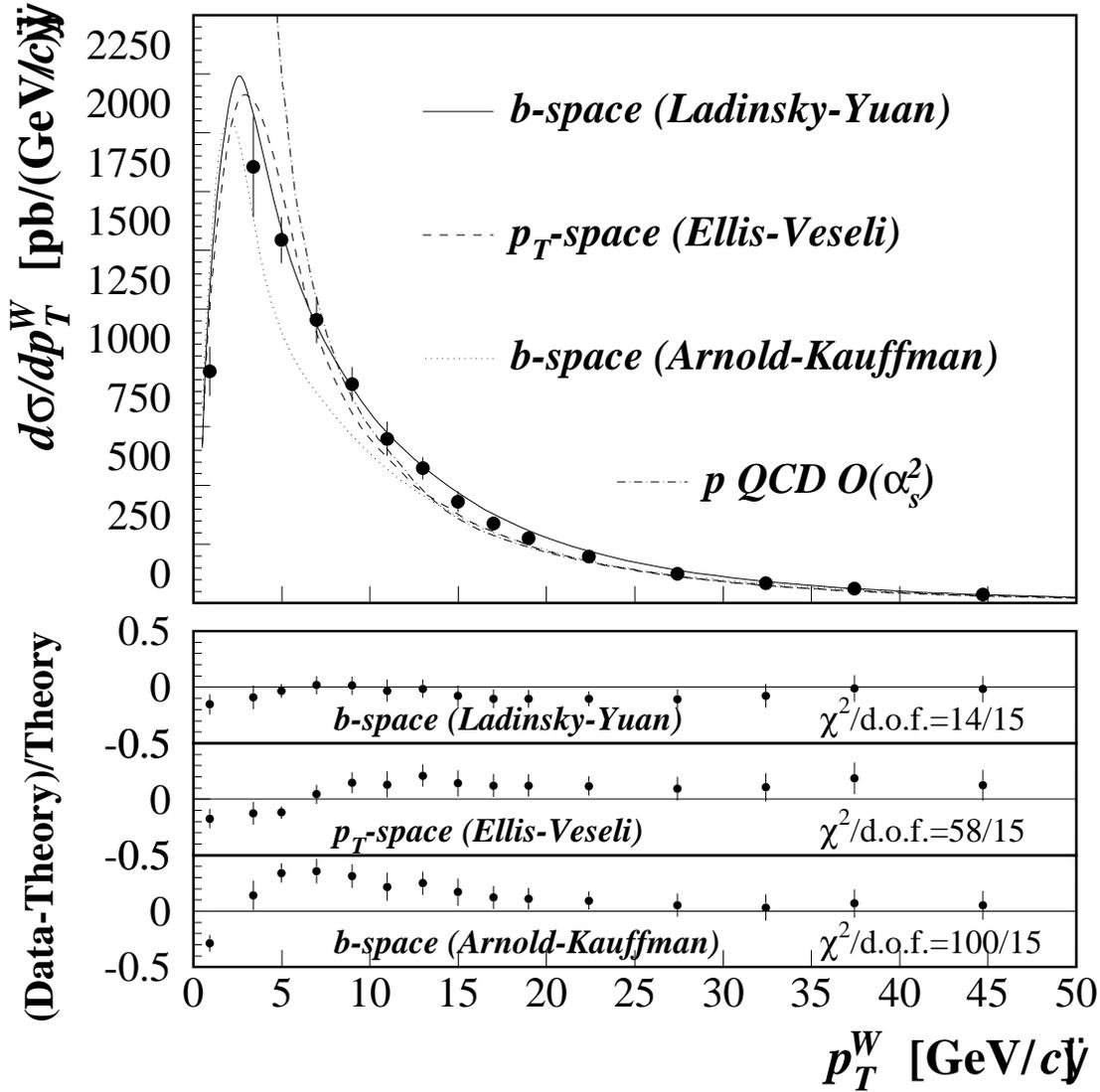,width=15cm}}
\caption{Differential cross section for $W$ boson production 
compared to three resummation calculations and to the fixed-order calculation.
Uncertainties on data include both statistical and systematical contributions 
(other than an overall normalization uncertainty in integrated luminosity).
Also shown are the fractional differences (Data--Theory)/Theory 
between data and the resummed predictions.}
\label{fig:ratio}
\end{figure}

Figure~\ref{fig:ratio} shows a comparison of 
the differential cross section for $W$ boson production, 
assuming $\mbox{B}(W\rightarrow e\nu)=0.111$, 
to the fixed-order perturbative calculation 
and to three different resummation calculations 
in the low $p_T^W$ region.
The parametrizations of the nonperturbative region are from 
Arnold-Kauffman~\cite{Arnold:1991yk} and Ladinsky-Yuan~\cite{Ladinsky:1994zn} 
in $b$-space, and Ellis-Veseli~\cite{Ellis:1998ii} in $p_T$-space. 
The disagreement between the data and the fixed-order prediction at 
low values of $p_T^W$ confirms the presence of contributions from soft gluon emission, 
which are accounted for in the resummation formalisms.
The fractional differences $(\mbox{Data} - \mbox{Theory})/\mbox{Theory}$ 
are also shown in Fig.~\ref{fig:ratio} for each of the three resummation predictions.
Although the $\chi^2$ for the Ellis-Veseli and Arnold-Kauffman prescriptions
are not as good as for Ladinsky-Yuan, 
the flexibility in parameter space and in the form of the nonperturbative function
in all three resummed models is such that a good description of our measurement 
can be achieved~\cite{Abbott:2000wk,Ellis:1997sc}.

\begin{figure}[p]
\centerline{\psfig{figure=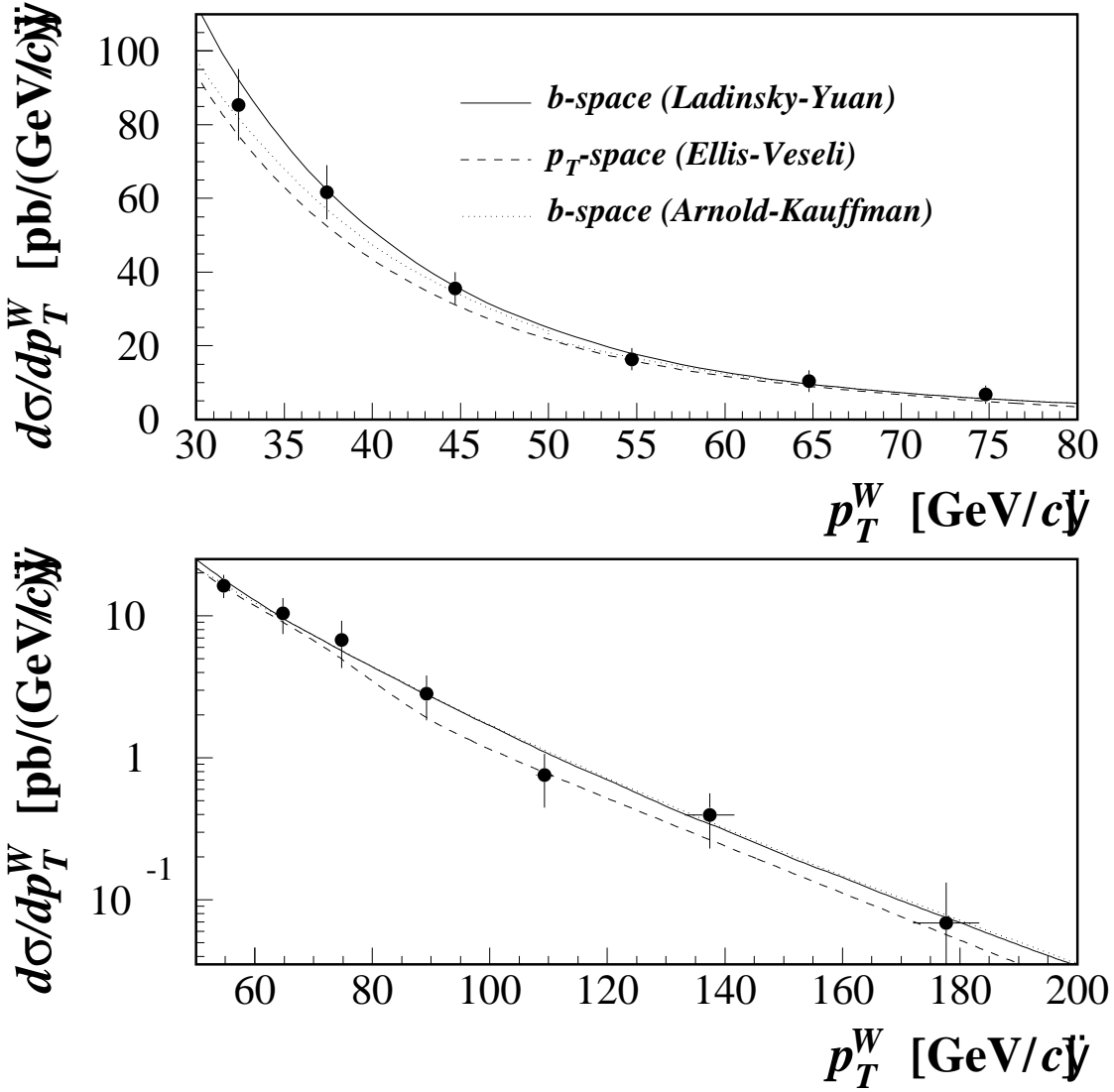,width=15cm}}
\caption{Differential cross section for $W$ boson production 
in the intermediate and high $p_T^W$ regions. 
Uncertainties on data include both statistical and systematical contributions
(other than an overall normalization uncertainty in integrated luminosity).}
\label{fig:intermediate}
\end{figure}

Figure~\ref{fig:intermediate} shows the differential cross section for $W$ boson production 
in the intermediate and high $p_T^W$ regions.
The calculation by Ladinsky-Yuan~\cite{Ladinsky:1994zn} 
specifies a matching prescription which provides a
smooth transition between the resummed and the fixed-order perturbative results to
${\mathcal O}(\alpha_s^2)$.
The $p_T$-space result by Ellis-Veseli~\cite{Ellis:1998ii} contains only 
the ${\mathcal O}(\alpha_s)$ finite part 
and an ${\mathcal O}(\alpha_s^2)$ Sudakov form factor.
Hence, there is still a residual unmatched higher-order effect 
present in $d\sigma/dp_T^W$ in the large $p_T^W$ region, 
where the cancellation of the different parts is quite delicate.
The $b$-space prediction by Arnold-Kauffman~\cite{Arnold:1991yk} 
uses the matched result below
$p_T^W = 50\ \mbox{GeV}/c$ and the pure conventional perturbative 
${\mathcal O}(\alpha_s^2)$ result above.
We observe good agreement with the theoretical predictions for 
intermediate and high values of $p_T^W$,
which probes effects of fixed-order QCD.

In summary, we have used data taken with the D\O\ detector in $p\bar{p}$ collisions at 
$\sqrt{s} = 1.8\ \mbox{TeV}$ to 
measure the cross section for $W\rightarrow e\nu$ events as a function of $p_T^W$. 
The combined QCD perturbative and resummed predictions
are in agreement with the fully corrected $p_T$ spectrum of $W$ boson production
in the kinematic range of the measurement.

We thank the staffs at Fermilab and collaborating institutions, 
and acknowledge support from the 
Department of Energy and National Science Foundation (USA),  
Commissariat  \` a L'Energie Atomique and 
CNRS/Institut National de Physique Nucl\'eaire et 
de Physique des Particules (France), 
Ministry for Science and Technology and Ministry for Atomic 
   Energy (Russia),
CAPES and CNPq (Brazil),
Departments of Atomic Energy and Science and Education (India),
Colciencias (Colombia),
CONACyT (Mexico),
Ministry of Education and KOSEF (Korea),
CONICET and UBACyT (Argentina),
The Foundation for Fundamental Research on Matter (The Netherlands),
PPARC (United Kingdom),
A.P. Sloan Foundation,
and the A. von Humboldt Foundation.

\newpage
\begin{table}[t]
\caption{Summary of the measurement of the $p_T$ distribution of the $W$ boson.
The nominal $p_T^W$ is where the predicted function equals its mean value over the bin.
The quantity $d\sigma(W\rightarrow e\nu)/dp_T^W$ corresponds to the 
differential cross section in each bin of $p_T^W$ for $W\rightarrow e\nu$ production.
Systematic uncertainties do not include an 
overall $4.4 \%$ normalization uncertainty in integrated luminosity.}
\begin{center}
\begin{tabular}{@{ }r@{\hspace{1em}}r@{--}l@{\hspace{1em}}c@{$\pm$}c@{$\pm$}c} 
\multicolumn{1}{c@{ }}{ } & 
\multicolumn{3}{c@{\hspace{1em}}}{ } &
\multicolumn{1}{c@{\hspace{1em}}}{Statistical } & Systematic \\
\multicolumn{1}{c@{ }}{\raisebox{1.5ex}[0pt]{$p_T^W$}} & 
\multicolumn{2}{c@{\hspace{1em}}}{\raisebox{1.5ex}[0pt]{$p_T^W$ bin}} &
\multicolumn{1}{c@{\hspace{1em}}}{\makebox[15pt][l]{\raisebox{1.5ex}[0pt]{$\frac{\displaystyle d\sigma}{\displaystyle dp_{\scriptscriptstyle T}^{\scriptscriptstyle W}}$}}\raisebox{2.6ex}[0pt]{$\scriptstyle (W\rightarrow e\nu)$}} & 
\multicolumn{1}{c@{\hspace{1em}}}{uncertainty} & uncertainty \\[1.5ex]
{\footnotesize GeV$/c$} & \multicolumn{2}{c@{\hspace{1em}}}{{\footnotesize GeV$/c$}} & 
\multicolumn{1}{c@{\hspace{1em}}}{{\footnotesize pb/(GeV$/c$)}} & 
\multicolumn{1}{c@{\hspace{1em}}}{{\footnotesize pb/(GeV$/c$)}} & {\footnotesize pb/(GeV$/c$)} \\[0.6ex] \hline
  0.92 &   0 &   2 & 109.37   & 4.60   & 10.64   \\
  3.40 &   2 &   4 & 205.91   & 6.84   & 22.80   \\
  4.97 &   4 &   6 & 171.28   & 5.64   &  9.16   \\
  6.98 &   6 &   8 & 133.62   & 4.65   &  9.81   \\
  8.98 &   8 &  10 & 103.30   & 4.03   &  7.17   \\
 10.98 &  10 &  12 &  77.58   & 3.47   &  7.15   \\
 12.98 &  12 &  14 &  63.66   & 3.21   &  4.18   \\
 14.98 &  14 &  16 &  47.88   & 2.77   &  4.03   \\
 16.98 &  16 &  18 &  37.72   & 2.43   &  2.50   \\
 18.98 &  18 &  20 &  30.65   & 2.21   &  1.60   \\
 22.40 &  20 &  25 &  22.02   & 1.23   &  1.11   \\
 27.41 &  25 &  30 &  13.94   & 0.93   &  0.98   \\
 32.42 &  30 &  35 &   9.47   & 0.73   &  0.79   \\
 37.42 &  35 &  40 &   6.84   & 0.63   &  0.52   \\
 44.70 &  40 &  50 &   3.95   & 0.36   &  0.31   \\
 54.72 &  50 &  60 &   1.81   & 0.24   &  0.23   \\
 64.77 &  60 &  70 &   1.15   & 0.21   &  0.25   \\
 74.79 &  70 &  80 &   0.75   & 0.18   &  0.21   \\
 89.21 &  80 & 100 &   0.313  & 0.059  &  0.091  \\
109.27 & 100 & 120 &   0.084  & 0.029  &  0.018  \\
137.40 & 120 & 160 &   0.044  & 0.012  &  0.014  \\
177.64 & 160 & 200 &   0.0077 & 0.0054 &  0.0045 \\ 
\end{tabular}
\end{center}
\label{tab:ptw}
\end{table}

\newpage

\newpage

\newpage
\end{document}